\documentclass[11pt,a4paper]{article}
\usepackage{a4wide}
\usepackage{amsmath,amssymb}
\usepackage[mathscr]{eucal}
\newtheorem{theorem}{Theorem}
\newtheorem{definition}{Definition}
\newtheorem{lemma}{Lemma}
\newtheorem{proposition}{Proposition}
\begin{document}
\title{Self-adjoint Lyapunov variables, temporal ordering and\\ 
irreversible representations of Schr\"odinger evolution}
\author{}
\date{}
\maketitle
\begin{description}
\item{$\ \ \ \ \ $} {\bf Y. Strauss \footnote{E-mail: ystrauss@math.bgu.ac.il \par \hskip1.5cm  yossef.strauss@gmail.com}}\\
              {\emph{\footnotesize{Einstein Institute of Mathematics, The Hebrew University of Jerusalem, Jerusalem, 91904, Israel}}}
\end{description}
\smallskip
\begin{abstract}
In non relativistic quantum mechanics time enters as a parameter in the Schr\"odinger equation. However, there are various 
situations where the need arises to view time as a dynamical variable. In this paper we consider the dynamical role 
of time through the construction of a Lyapunov variable - i.e., a self-adjoint quantum observable 
whose expectation value varies monotonically as time increases. It is shown, in a constructive way, that a certain class of 
models admit a Lyapunov variable and that the existence of a Lyapunov variable implies the existence of 
a transformation mapping the original quantum mechanical problem to an equivalent 
irreversible representation. In addition, it is proved that in the irreversible representation there exists a natural time 
ordering observable splitting the Hilbert space at each $t>0$ into past and future subspaces.
\end{abstract}
\section{Introduction}
\label{introduction}
In standard, non-relativistic, quantum mechanics time enters as a parameter in the Schr\"odinger equation governing the evolution 
of a quantum system and may be thought of as a label for the various states obtained by the system in the course of its 
evolution. As such, time is not a dynamical variable and is considered to be an external 
variable measured by laboratory clocks which are not dynamically related to the objects under investigation. However, in 
various contexts and for various reasons the need arises to extend beyond this traditional role of time and to view it in 
some sense as a \emph{dynamical variable}. In fact, the question of the dynamical 
nature of time in quantum mechanics emerged already in its early days with regard to the validity and meaning of the 
time-energy uncertainty relation (see e.g. the review by P. Busch in Ref. 1 and references therein).
In particular, Pauli${}^2$ has shown that, for a semi-bounded Hamiltonian $H$, 
it is impossible to find a self-adjoint operator $T$ such that $H$ and $T$ satisfy the canonical commutation relations (CCR) 
$[H,T]=i$ and constitute an imprimitivity system, i.e., it is impossible to find a pair $H$, $T$ with $H$ self-adjoint and 
semi-bounded and $T$ self-adjoint, such that the covariance relations
\begin{subequations}
\begin{equation}
\label{imprimitivity_sys_a_eqn}
 e^{i H t} T e^{-i H t} = T+t I
\end{equation}
\begin{equation}
\label{imprimitivity_sys_b_eqn} 
 e^{i T \xi} H e^{-i T\xi} = H-\xi I
\end{equation}
\end{subequations}
are both satisfied. Pauli's theorem has, erroneously, been taken by many to imply that a definition of a self-adjoint operator $T$
satisfying the CCR with $H$ is categorically impossible. In fact, the CCR is weaker than the existence of an imprimitivity system,
i.e., the existence of an imprimitivity system as in Eqns. (\ref{imprimitivity_sys_a_eqn}), 
(\ref{imprimitivity_sys_b_eqn}) implies the validity of the CCR $[H,T]=i$, but in
many cases it is possible to find a self-adjoint $T$ such that $H$ and $T$ satisfy the CCR but do not satisfy the covariance 
relations in Eqns. (\ref{imprimitivity_sys_a_eqn}), (\ref{imprimitivity_sys_b_eqn})${}^{3,4,5}$. Whenever an operator $T$ 
satisfying the CCR with the Hamiltonian $H$ can be found it is referred to as a \emph{canonical time operator}. We note that
there are cases, such as for the free particle, where the canonical time operator is a maximally symmetric operator with no self-adjoint extensions. 
\par The need for a dynamical point of view with regard to time in quantum mechanics is distinctly emphasized in experiments 
aimed at measuring the time of occurrence of events, for example, experiments involving the measurement of the time of arrival of
a particle at a detector. In such experiments, where the temporal statistics of the occurrence of events is being measured, time is 
to be considered as an observable. Taking into account the fact that in many cases a given Hamiltonian may not admit a 
self-adjoint canonical time operator, or does not admit a canonical time operator at all, an operational approach to 
experiments involving time measurments has been adopted in which a time observable is defined in terms of a corresponding 
positive operator valued measure (POVM)${}^{6,7,8,9,10}$. 
Using such a POVM one can define a time operator which is, in general, a maximally symmetric, non self-adjoint operator. 
\par In the present paper we take, in a sense, a relatively conservative point of view with regard to the problem of time as a 
dynamical variable in quantum mechanics. We search not for a time observable, neither in the form of a standard, self-adjoint 
quantum observable nor one defined in terms of a more general POVM, but instead seek a definition of a self-adjoint 
observable satisfying a minimal set of requirements which enable it to serve as a marker for the flow of time for the system being 
studied. Our starting point is the simple observation that every non-stationary dynamical variable provides an indication for the 
flow of time in a quantum system via the change of its expectation value. This may be regarded as the simplest manifestation of a 
dynamical point of view of time. Moreover, by marking the passage of time not by external laboratory clocks, but through the 
change in expectation value of a dynamical variable intrinsic to the system, one obtains a notion of an \emph
{intrinsic time} of a system${}^{1}$. However, the use of a non-stationary observable as a marker for the flow of time in 
a quantum system immediately encounters a fundamental problem of generality which we consider presently.
\par Let $\mathcal H$ be a Hilbert space corresponding to a given quantum mechanical system and let $H$ be a self-adjoint
Hamiltonian generating the evolution of the system. Let $\varphi(t)=U(t)\varphi=\exp(-i Ht)\varphi$ be the state of the system at 
time $t$ corresponding to an initial state $\varphi(0)=\varphi\in\mathcal H$. Define the \emph{trajectory} 
$\Phi_\varphi$ corresponding to an initial state $\varphi\in\mathcal H$ to be
\begin{equation*}
 \Phi_\varphi:=\{U(t)\varphi\}_{t\in\mathbb R^+}=\{\varphi(t)\}_{t\in\mathbb R^+}\,,
\end{equation*}
i.e., $\Phi_\varphi$ is the set of states reached in the course of the evolution of the system from an initial state 
$\varphi$. We shall call a trajectory $\Phi_\varphi$ \emph{recurring} if there exist two times $0\leq t_1<t_2$ such that the states of the system at $t_1$ and at $t_2$
differ only by a global phase, i.e, there exists some $\alpha\in\mathbb R$ and $t_1,t_2\in\mathbb R^+$,
with $t_1\neq t_2$, such that $\varphi(t_1)=e^{i\alpha}\varphi(t_2)$. A trajectory $\Phi_\varphi$ is \emph{non-recurring} if the condition 
$\varphi(t_1)=e^{i\alpha}\varphi(t_2)$ is satisfied only for $t_1=t_2$ and $\alpha=2\pi k$, $k\in\mathbb Z$.
Next, let $\mathfrak B(\mathcal H)$ denote the space of bounded linear operators on $\mathcal H$. We define the notion of a (forward) Lyapunov variable as follows: 
\begin{definition}[(forward) Lyapunov variable]
Let $M\in\mathfrak B(\mathcal H)$ be a self-adjoint operator on $\mathcal H$.
Let $\Phi_\varphi$ be an arbitrary trajectory corresponding to a chosen initial state $\varphi\in\mathcal H$. 
Denote by $M(\Phi_\varphi)=\{ (\psi,M\psi)\mid \psi\in\Phi_{\varphi}\}$ the collection of all expectation values of $M$
for states in $\Phi_\varphi$ (up to a multiplicative normalization constant). Then $M$ is a (forward) Lyapunov variable 
if the mapping $\tau_{M,\varphi}: \mathbb R^+\mapsto M(\Phi_\varphi)$ defined by
\begin{equation*}
 \tau_{M,\varphi}(t)=(\varphi(t), M\varphi(t))
\end{equation*}
is bijective and monotonically decreasing for all non-recurring trajectories. \hfill$\square$
\end{definition}
\par{\bf Remark:} The requirement that $\tau_{M,\varphi}$ be monotonically decreasing is made purely for the sake of convenience.
If we require that $\tau_{M,\varphi}$ is monotonically increasing we obtain a perefectly valid definition of a forward Lyapunov 
variable. 
\par{}
\bigskip
If $M$ is a Lyapunov variable as defined above then we are able to find the temporal ordering of the states
belonging to a trajectory $\Phi_\varphi$ in an obvious way according to the natural ordering of the expectation values in 
$M(\Phi_\varphi)$. In terms of the definition of a forward Lyapunov variable the aformentioned problem of generality 
is stated in a simple way: It may happen that for a given non-stationary observable $A$ and a given trajectory $\Phi_\varphi$ the function 
$\mathbb R^+\ni t\to (\varphi(t), A\,\varphi(t))\in\mathbb R$ is monotonically decreasing. It is then 
possible to find the temporal ordering of the states in $\Phi_\varphi$ according to the expectation values $(\psi, A\psi)$ where
$\psi\in\Phi_\varphi$. However, unless $A$ is a Lyapunov variable, this property is not general, i.e., there are other trajectories for 
which the sequence of expectation values of $A$ is not monotonic or even monotonically increasing. In this case there will be 
trajectories for which the proper temporal ordering of states cannot be deciphered in a simple way from the knowledge of the 
expectation values of $A$.
\par The existence of a Lyapunov variable introduces a global temporal ordering in the Hilbert space. Suppose that a forward 
Lyapunov variable $M_F$ can be defined for a given quantum mechanical problem and, for simplicity, assume that all 
trajectories are non-recurring. In the Hilbert space $\mathcal H$ for the problem we define the subsets
\begin{equation*}
 \mathcal F_m=\{\psi\in\mathcal H \mid \Vert\psi\Vert^{-2}(\psi, M_F\,\psi)\leq m \}\,.
\end{equation*}   
Set $m_l=\inf_{\Vert\psi\Vert=1}(\psi,M_F\,\psi)$ and $m_u=\sup_{\Vert\psi\Vert=1}(\psi,M_F\,\psi)$. Then, for 
$m_l\leq m_2\leq m_1\leq m_u$ we necessarily have
\begin{equation*}
 \mathcal F_{m_2}\subseteq\mathcal F_{m_1}\,.
\end{equation*}
Hence the class of sets $\{\mathcal F_m\}_{m_l\leq m\leq m_u}$ is ordered according to the values of $m$. This ordering 
amounts to a temporal ordering in $\mathcal H$ in the sense that if for two states $\psi_1,\psi_2\in\mathcal H$ we have 
$\psi_2\in\mathcal F_m$ and $\psi_1\in (\mathcal H\backslash\mathcal F_m)$ and if there exists a trajectory $\Phi_\varphi$, 
starting at an initial state $\varphi\in\mathcal H$, such that $\psi_1,\psi_2\in\Phi_\varphi$ and $\psi_1=\varphi(t_1)$, 
$\psi_2=\varphi(t_2)$, then we must have $t_2\geq t_1$. Put simply, states in $\mathcal F_m$ can lie in the future of states 
in $\mathcal H\backslash\mathcal F_m$ but not vice versa. 
\par In Theorem \ref{exists_lyapunov_var_thm} it is shown that, essentially under the condition that the spectrum of the 
Hamiltonian $H$ satisfies $\sigma_{ac}(H)=\mathbb R^+$, it is possible to define on the subspace $\mathcal H_{ac}\subset\mathcal H$ corresponding to the absolutely continuous spectrum of $H$, a forward Lyapunov variable which will be 
denoted by $M_F$. The proof of existence and properties of $M_F$ utilizes some basic tools from the theory of Hardy spaces${}^{11,12,13}$. 
However, beyond the existence of $M_F$, it is of no less significance to note the fact that an extension of
the mathematical constructions and techniques used in the proof of Theorem \ref{exists_lyapunov_var_thm} 
imply some interesting consequences to the existence of the Lyapunov variable. 
Specifically, if $\{U(t)\}_{t\in\mathbb R}$, where $U(t)=\exp(-i Ht)$, is the Schr\"odinger evolution group of the quantum 
system then it is possible to define a contractive one-parameter continuous semigroup $\{Z(t)\}_{t\in\mathbb R^+}$ satisfying
\begin{equation*}
 \Vert Z(t_2)\psi\Vert\leq \Vert Z(t_1)\psi\Vert,\qquad t_2\geq t_1,\quad \psi\in\mathcal H_{ac}
\end{equation*}
and
\begin{equation*}
 s-\lim_{t\to\infty} Z(t)=0\,,
\end{equation*}
and a non-unitary mapping $\Lambda_F$, called a $\Lambda$-transformation, with the intertwining property
\begin{equation*}
 \Lambda_F U(t)=Z(t)\Lambda_F,\quad t\geq 0\,.
\end{equation*}
One is then able to obtain an \emph{irreversible representation} of the dynamics. This is done in Section 
\ref{main_results_sec} below.
In the irreversible representation the dynamics of the system is unidirectional in time and is given in terms of the semigroup 
$\{Z(t)\}_{t\in\mathbb R^+}$. 
\par It is an interesting fact that in the irreversible representation of the dynamics it is possible to find a positive 
semibounded operator $T$ on $\mathcal H_{ac}$ that can be intepreted as a natural time ordering observable for the evolution of 
the system. The exact nature of this observable and the main theorem concerned with its existence is discussed 
in Section \ref{main_results_sec}. Of course, this operator is not a time operator in the sense of 
Eqns. (\ref{imprimitivity_sys_a_eqn}), (\ref{imprimitivity_sys_b_eqn}), nor is it a cannonical time operator in the weaker
sense that $T$ satisfies the CCR with the Hamiltonian $H$.
\par The idea of finding, once the existence of a Lyapunov variable is established, a transformation of the unitary quantum 
evolution into an irreversible representation of the quantum dynamics in terms of semigroup evolution follows directly from a framework
developed by B. Misra, I. Prigogine and M. Courbage (MPC) and various other contributors (see for example 
Refs. 14, 15, 16, 17, 18, 19 and references therein) who, starting in the late 1970's and through the following decades, 
developed a theory of classical and quantum microscopic irreversibility. A central notion in the MPC theory
is that of a non-equilibrium entropy associated with the existence of a Lyapounov variable. The mathematical constructions
and results of the present paper may have implications within the MPC formalism, but it should be emphasized that in the 
present paper we do not work within the MPC framework and we do not assign the Lyapunov variable $M_F$ the meaning of 
non-equillibrium entropy. Indeed, without further modifications the operator $M_F$ introduced below does not qualify as a 
good representer of non-equilibrium entropy in the MPC theory as it does not satisfy some of the conditions rquired by MPC 
from such an object. 
\par The main theorems proved in this paper, concerning the 
existence of a Lyapunov variable, the transformation to an irreversible representation and existence of a natural self-adjoint 
time ordering observable in the irreversible representation, are stated in Section \ref{main_results_sec}. The proofs of these 
theorems are provided in Section \ref{proofs_sec}. A short summary is provided in Section \ref{summary}.
\section{Main theorems and results}
\label{main_results_sec}
\par We consider quantum mechanical dynamical problems for which the Hamiltonian $H$ generating the Schr\"odinger evolution satisfies the
condition that $\sigma_{ac}(H)=\mathbb R^+$ . We shall assume that in the energy represention, i.e. the spectral representation for the 
Hamiltonian $H$, the subspace $\mathcal H_{ac}\subseteq\mathcal H$, corresponding to the absolutely continuous spectrum of $H$, is represented
by the function space $L^2(\mathbb R^+;\mathcal K)$ of $\mathcal K$ valued functions with $\mathcal K$ a seperable Hilbert space. The discussion in 
the present paper is essentially based on the spectral representation for the Hamiltonian and does not depend on the details of the physical model 
considered. Thus we take as our basic setting the function space $L^2(\mathbb R^+;\mathcal K)$ and on this space we consider the evolution group 
$\{u_+(t)\}_{t\in\mathbb R}$ defined in Eq. (\ref{spec_schrodinger_evolution_eqn}) below whose generator is the operator of multiplication by the 
independent variable on $L^2(\mathbb R^+;\mathcal K)$.
\par The three theorems in this section, and the discussion accompanying them, provide the main results of the present
paper. In order to state these theorems we need a few basic facts concerning Hardy spaces of vector valued functions.
Denote by $\mathbb C^+$ the upper half of the complex plane and let $\mathcal K$ be a seperable Hilbert space. The Hardy 
space $\mathcal H^2(\mathbb C^+;\mathcal K)$ of $\mathcal K$ valued functions defined on $\mathbb C^+$ consists of 
$\mathcal K$ valued functions analytic in $\mathbb C^+$ and satisfying the condition that for any 
$f\in\mathcal H^2(\mathbb C^+;\mathcal K)$ there exists a constant $0<C_f<\infty$ such that 
\begin{equation*}
 \sup_{y>0}\int_{-\infty}^\infty dx\,\Vert f(x+iy)\Vert_{\mathcal K}^2\leq C_f\,.
\end{equation*}
In a similar manner the Hardy space $\mathcal H^2(\mathbb C^-;\mathcal K)$ consists of functions analytic in the lower 
half-plane $\mathbb C^-$ and satisfying the condition that for any $g\in\mathcal H^2(\mathbb C^-;\mathcal K)$ there exists a 
constant $0<C_g<\infty$ such that 
\begin{equation*} 
 \sup_{y>0}\int_{-\infty}^\infty dx\,\Vert g(x+iy)\Vert_{\mathcal K}^2\leq C_g\,.
\end{equation*}
Hardy space functions have non-tangential boundary values a.e. on $\mathbb R$. In particular, if $L^2(\mathbb R;\mathcal K)$ is
the Hilbert space of $\mathcal K$ valued $L^2$ functions defined on $\mathbb R$, then for any function 
$f\in\mathcal H^2(\mathbb C^+;\mathcal K)$ there exists a function $\tilde f\in L^2(\mathbb R;\mathcal K)$ such that a.e. on 
$\mathbb R$ we have
\begin{equation*}
 \lim_{y\to 0^+}\Vert f(x+iy)-\tilde f(x)\Vert_{\mathcal K}=0,\qquad x\in\mathbb R\,.
\end{equation*}
A similar limit from below the real axis holds for functions in $\mathcal H^2(\mathbb C^-;\mathcal K)$. In fact,
$\mathcal H^2(\mathbb C^\pm;\mathcal K)$ are Hilbert spaces with scalar product given by
\begin{multline*}
 (f,g)_{\mathcal H^2(\mathbb C^\pm;\mathcal K)}=\lim_{y\to 0^+}
 \int_{-\infty}^\infty dx\,(f(x\pm iy),g(x\pm iy))_{\mathcal K}=\\
 =\int_{-\infty}^\infty dx\,(\tilde f(x),\tilde g(x))_{\mathcal K},\quad f,g\in\mathcal H^2(\mathbb C^\pm;\mathcal K),
\end{multline*}
where $\tilde f,\tilde g$ are, respectively, the boundary value functions of $f$ and $g$. The spaces of boundary value 
functions on $\mathbb R$ of functions in $\mathcal H^2(\mathbb C^\pm;\mathcal K)$ are then Hilbert spaces isomorphic to
$\mathcal H^2(\mathbb C^\pm;\mathcal K)$ which we denote by $\mathcal H^2_\pm(\mathbb R;\mathcal K)$. The Hardy spaces
$\mathcal H^2_\pm(\mathbb R;\mathcal K)$ are subspaces of $L^2(\mathbb R;\mathcal K)$. Moreover, we have
\begin{equation*} 
 L^2(\mathbb R;\mathcal K)=\mathcal H^2_+(\mathbb R;\mathcal K)\oplus\mathcal H^2_-(\mathbb R;\mathcal K)\,.
\end{equation*}
We shall denote the orthogonal projection of $L^2(\mathbb R;\mathcal K)$ onto $\mathcal H^2_+(\mathbb R;\mathcal K)$ by 
$P_+$ and the orthogonal projection of $L^2(\mathbb R;\mathcal K)$ onto $\mathcal H^2_-(\mathbb R;\mathcal K)$ by 
$P_-$. 
\par Denote by $L^2(\mathbb R^+;\mathcal K)$ the subspace of $L^2(\mathbb R;\mathcal K)$ of functions supported on
$\mathbb R^+$ and let $L^2(\mathbb R^-;\mathcal K)$ be the subspace of $L^2(\mathbb R;\mathcal K)$ containing functions 
supported on $\mathbb R^-$. Then we have another orthogonal decomposition of $L^2(\mathbb R;\mathcal K)$, i.e.,
\begin{equation*} 
 L^2(\mathbb R;\mathcal K)=L^2(\mathbb R^+;\mathcal K)\oplus L^2(\mathbb R^-;\mathcal K)\,.
\end{equation*}
We denote the orhotgonal projection of $L^2(\mathbb R;\mathcal K)$ onto $L^2(\mathbb R^+;\mathcal K)$ by $P_{\mathbb R^+}$.
Similarly, the orhotgonal projection of $L^2(\mathbb R;\mathcal K)$ onto $L^2(\mathbb R^-;\mathcal K)$ will be denoted by 
$P_{\mathbb R^-}$. On $L^2(\mathbb R^+;\mathcal K)$ define a family $\{u_+(t)\}_{t\in\mathbb R}$ of unitary multiplicative
operators $u_+(t): L^2(\mathbb R^+;\mathcal K)\mapsto L^2(\mathbb R^+;\mathcal K)$ by
\begin{equation}
\label{spec_schrodinger_evolution_eqn}
 [u_+(t)f](E)=e^{-iEt}f(E),\qquad f\in L^2(\mathbb R^+;\mathcal K),\ \ E\in\mathbb R^+\,.
\end{equation}
The family $\{u_+(t)\}_{t\in\mathbb R}$ forms a continuous one parameter unitary group on $L^2(\mathbb R^+;\mathcal K)$. This 
group represents the Schr\"odinger evolution in the energy representation. 
\par We can now state our first theorem concerning the construction of Lyapunov variables for Schr\"odinger evolution
\begin{theorem}
\label{exists_lyapunov_var_thm}
Let $M_F: L^2(\mathbb R^+;\mathcal K)\mapsto L^2(\mathbb R^+;\mathcal K)$ be the operator defined by
\begin{equation}
\label{M_F_expression_eqn}
 M_F:=(P_{\mathbb R^+}P_+P_{\mathbb R+})\vert_{L^2(\mathbb R^+;\mathcal K)},
\end{equation}
i.e., $M_F$ is the restriction of the operator $P_{\mathbb R^+}P_+P_{\mathbb R^+}$, defined on $L^2(\mathbb R;\mathcal K)$, to the subspace 
$L^2(\mathbb R^+;\mathcal K)$. Then $M_F$ is a self-adjoint, contractive, injective and non-negative operator on $L^2(\mathbb R^+;\mathcal K)$ such 
that $Ran\,M_F$ is dense in $L^2(\mathbb R^+;\mathcal K)$ and $M_F$ is a Lyapunov variable for the Schr\"odinger
evolution on $L^2(\mathbb R^+;\mathcal K)$ in the forward direction, i.e, for every $\psi\in L^2(\mathbb R^+;\mathcal K)$ we 
have
\begin{equation}
\label{M_F_lyapounov_var_eqn}
 (\psi_{t_2},M_F\psi_{t_2})\leq(\psi_{t_1},M_F\psi_{t_1}),\qquad t_2\geq t_1\geq 0,\quad \psi_t=u_+(t)\psi
\end{equation}
and, moreover,
\begin{equation}
\label{M_F_to_zero_eqn}
 \lim_{t\to\infty}(\psi_t,M_F\psi_t)=0\,.
\end{equation}
\hfill$\square$
\end{theorem}
\par Following the proof of the existence of the Lyapunov variable $M_F$ we can proceed as in the MPC theory and obtain 
a non-unitary $\Lambda$-transformation via the definition $\Lambda_F:=M_F^{1/2}$, where $M_F^{1/2}$ is the positive square root of $M_F$. We then have the 
following theorem
\begin{theorem}
\label{lambda_transform_thm}
Let $\Lambda_F:=M_F^{1/2}$. Then $\Lambda_F: L^2(\mathbb R^+;\mathcal K)\mapsto L^2(\mathbb R^+;\mathcal K)$ is positive, 
contractive and quasi-affine, i.e., it is a positive, contractive, injective operator such that $Ran\,\Lambda_F$ is dense in 
$L^2(\mathbb R^+;\mathcal K)$. Furthermore, there exists a continuous, strongly contractive, one parameter semigroup 
$\{Z(t)\}_{t\in\mathbb R^+}$, with $Z(t): L^2(\mathbb R^+;\mathcal K)\mapsto L^2(\mathbb R^+;\mathcal K)$, such that for 
every $\psi\in L^2(\mathbb R^+;\mathcal K)$ we have
\begin{equation}
\label{Z_semigroup_prop_eqn}
 \Vert Z(t_2)\psi\Vert\leq \Vert Z(t_1)\psi\Vert,\ \ t_2\geq t_1\geq 0 
\end{equation}
and
\begin{equation*}
 s-\lim_{t\to\infty} Z(t)=0
\end{equation*}
and the following intertwining relation holds
\begin{equation}
\label{Z_intertwining_eqn}
 \Lambda_F u_+(t)=Z(t)\Lambda_F,\qquad t\geq 0\,.
\end{equation}
\hfill$\square$
\end{theorem}
Taking the adjoint of Eq. (\ref{Z_intertwining_eqn}) we obtain the intertwining relation
\begin{equation}
\label{Z_intertwining_adj_eqn}
 u_+(-t)\Lambda_F=\Lambda_F Z^*(t),\quad t\geq 0\,.
\end{equation}
Let $\mathscr L(\mathcal H)$ denote the set of linear operators on a Hilbert space $\mathcal H$. 
Let $X\in\mathscr L(L^2(\mathbb R^+;\mathcal K))$ be self-adjoint and set $X_{\Lambda_F}:=\Lambda_F^{-1}X\Lambda_F^{-1}$, so 
that $X=\Lambda_F X_{\Lambda_F}\Lambda_F$. Using Eqns. (\ref{Z_intertwining_eqn}) and (\ref{Z_intertwining_adj_eqn}) we obtain
\begin{equation}
\label{heisenberg_to_semigroup_eqn}
 u_+(-t)Xu_+(t)=
 u_+(-t)\Lambda_F X_{\Lambda_F}\Lambda_F u_+(t)=\Lambda_F Z^*(t) X_{\Lambda_F} Z(t)\Lambda_F,\quad t\geq 0\,.
\end{equation}
For $\varphi,\psi\in L^2(\mathbb R^+;\mathcal K)$ denote $\varphi_{\Lambda_F}=\Lambda_F\varphi$ and 
$\psi_{\Lambda_F}=\Lambda_F\psi$. Then Eq. (\ref{heisenberg_to_semigroup_eqn}) implies that
\begin{equation}
\label{QM_semigroup_rep_eqn}
 (\varphi,u_+(-t)X u_+(t)\psi)=(\varphi,\Lambda_FZ^*(t)X_{\Lambda_F}Z(t)\Lambda_F\psi)
 =(\varphi_{\Lambda_F},Z^*(t)X_{\Lambda_F}Z(t)\psi_{\Lambda_F}),\quad t\geq 0\,.
\end{equation}
Every relevant physical observable $X$ we shall consider will be assumed to have a representation in the form
$X=\Lambda_F X_{\Lambda_F}\Lambda_F$ for some self-adjoint $X_{\Lambda_F}\in\mathscr L(L^2(\mathbb R^+;\mathcal K))$. Then, 
since the left hand side of Eq. (\ref{QM_semigroup_rep_eqn}) corresponds to the original quantum mechanical problem, the 
right hand side of this equation constitutes a new representation of the original problem in terms of the correspondence
\begin{eqnarray*}
 \psi &\longrightarrow& \psi_{\Lambda_F}=\Lambda_F\psi\\
 u_+(t) &\longrightarrow& Z(t)=\Lambda_F\,U(t)\Lambda_F^{-1},\ \ t\geq 0\\
 X &\longrightarrow&\ \ \ X_{\Lambda_F}=\Lambda_F^{-1}X\Lambda_F^{-1}\,.
\end{eqnarray*}
Considering the fact that on the right hand side of Eq. (\ref{QM_semigroup_rep_eqn}) the dynamics is given in terms of the 
semigroup $\{Z(t)\}_{t\in\mathbb R^+}$  we may call the right hand side of 
Eq. (\ref{QM_semigroup_rep_eqn}) the \emph{irreversible representation} of the problem. The left hand side of that equation is
then the \emph{reversible representation} (or the standard representation). 
\par It is an interesting fact that in the irreversible representation of a quantum mechanical problem, as in the right hand side
of Eq. (\ref{QM_semigroup_rep_eqn}), one can find a self-adjoint operator $T$ with continuous spectrum $\sigma(T)=([0,\infty))$ 
such that for every $t\geq 0$ the spectral projections on the spectrum of $T$ naturally divide the Hilbert space $L^2(\mathbb R^+;\mathcal K)$
into a direct sum of a \emph{past subspace at time $t$} and a \emph{future subspace at time $t$}. Specifically, we have the following 
theorem:
\begin{theorem}
\label{T_observable_thm}
Let $\mathscr B(\mathbb R^+)$ be the Borel $\sigma$-algebra generated by open subsets of $\mathbb R^+$ and let
$\mathscr P(L^2(\mathbb R^+;\mathcal K))$ be the set of orthogonal projections in $L^2(\mathbb R^+;\mathcal K)$. There exists a semi-bounded, 
self-adjoint operator $T:\mathscr D(T)\mapsto L^2(\mathbb R^+;\mathcal K)$ defined on a dense domain $\mathscr D(T)\subset L^2(\mathbb R^+;\mathcal K)$
with continuous spectrum $\sigma(T)=[0,\infty)$ and corresponding spectral projection valued measure 
$\mu_T :\mathscr B(\mathbb R^+)\mapsto \mathscr P(L^2(\mathbb R^+;\mathcal K))$ such that for each $t\geq 0$ 
\begin{equation*}
 \mu_T([0,t])L^2(\mathbb R^+;\mathcal K)=[Z(t),Z^*(t)]L^2(\mathbb R^+;\mathcal K)=Ker\, Z(t),\qquad t\geq 0
\end{equation*}
and
\begin{equation*}
 \mu_T([t,\infty))L^2(\mathbb R^+;\mathcal K)=Z^*(t)Z(t)L^2(\mathbb R^+;\mathcal K)=(Ker\,Z(t))^\perp,\qquad t\geq 0\,.
\end{equation*}
In particular, for $0\leq t_1< t_2$ we have $Ker\,Z(t_1)\subset Ker\,Z(t_2)$. For $t=0$ we have $Ker\,Z(0)=\{0\}$ and finally
$\lim_{t\to\infty} Ker\,Z(t)=L^2(\mathbb R^+;\mathcal K)$. \hfill$\square$
\end{theorem}
Denote the orthogonal projection on $Ker\,Z(t)$ by $P_{t]}$ and the orthogonal projection on $(Ker\,Z(t))^\perp$ by $P_{[t}$\ . From
Theorem \ref{T_observable_thm} we have for $t\geq 0$
\begin{equation*}
 P_{t]}=[Z(t),Z^*(t)],\qquad P_{[t}=Z^*(t)Z(t),\ \ t\geq 0\,.
\end{equation*}
The projection $P_{t]}$ will be called below the \emph{projection on the past subspace at time $t$}. The projection $P_{[t}$ 
will be called the \emph{projection on the future subspace at time $t$}. In accordance we will call 
$L^2_{t]}(\mathbb R^+;\mathcal K):=Ran\,P_{t]}=Ker\,Z(t)$
the \emph{past subspace at time $t$} and $L^2_{[t}(\mathbb R^+;\mathcal K):=Ran\,P_{[t}=(Ker\,Z(t))^\perp$ the 
\emph{future subspace at time $t$}. The origin of the terminology used here can be found in Eq. (\ref{QM_semigroup_rep_eqn}). Using the notation for 
the projections on $Ker\,Z(t)$ and $(Ker\,Z(t))^\perp$ we observe that this equation may be written in the form
\begin{equation*}
 (\varphi,u_+(-t)X u_+(t)\psi)=(P_{[t}\,\varphi_{\Lambda_F},Z^*(t)X_{\Lambda_F}Z(t)P_{[t}\,\psi_{\Lambda_F}),\quad t\geq 0\,,
\end{equation*}
and denoting $\varphi_{\Lambda_F}^+(t):= P_{[t}\,\varphi_{\Lambda_F}=P_{[t}\Lambda_F\varphi$ and
$\psi_{\Lambda_F}^+(t):= P_{[t}\,\psi_{\Lambda_F}=P_{[t}\Lambda_F\psi$ we can write in short
\begin{equation}
\label{QM_semigroup_rep_B_eqn}
 (\varphi,u_+(-t)X u_+(t)\psi)=(\varphi^+_{\Lambda_F}(t),Z^*(t)X_{\Lambda_F}Z(t)\psi^+_{\Lambda_F}(t)),\quad t\geq 0\,.
\end{equation}
Note that in the irreversible representation on the right hand side of Eq. (\ref{QM_semigroup_rep_B_eqn}) only the projection
of $\varphi_{\Lambda_F}$ and $\psi_{\Lambda_F}$ on the future subspace $L^2_{[t}(\mathbb R^+;\mathcal K)$ at time $t$ is relevant for the 
calculation of all matrix elements and expectation values for times $t'\geq t\geq 0$. In other words, at time $t$ the subspace
$L^2_{t]}(\mathbb R^+;\mathcal K)=P_{t]}L^2(\mathbb R^+;\mathcal K) $ already belongs to the past and is irrelevant for calculations related to the 
future evolution of the system. We see that in the irreversible representation the spectral projections of the operator $T$ provide 
the time ordering of the evolution of the system. Following these observations it is natural to call $T$ a \emph{temporal 
ordering operator} for the irreversible representation. Note, in particular, that since $M_F=\Lambda_F^2$ we have 
$\Lambda_F^{-1}M_F\Lambda_F^{-1}=I$ and if we substitute this relation in Eq. (\ref{QM_semigroup_rep_eqn}) or 
Eq. (\ref{QM_semigroup_rep_B_eqn}) and take $\varphi=\psi$ we obtain
\begin{multline*}
 (\psi_t,M_F\,\psi_t)=(\psi,u_+(-t)M_Fu_+(t)\psi)=(\psi_{\Lambda_F},Z^*(t)Z(t)\psi_{\Lambda_F})=\\
 =(\psi_{\Lambda_F},P_{[t}\,\psi_{\Lambda_F})=(\psi_{\Lambda_F},\mu_T([t,\infty))\psi_{\Lambda_F}),\quad t\geq 0\,,
\end{multline*}
thus we have direct correspondence between the Lyapunov variable $M_F$ in the reversible representation of the problem
and the temporal ordering observable in the irreversible representation. 
\section{Proofs of Main results}
\label{proofs_sec}
\par The basic mechanism underlying the proofs of Theorem \ref{exists_lyapunov_var_thm} and 
Theorem \ref{lambda_transform_thm} is a fundamental intertwining relation, via a quasi-affine mapping, between the unitary 
Schr\"odinger evolution in the energy representation, i.e., the evolution group $\{u_+(t)\}_{t\in\mathbb R}$ on $L^2(\mathbb R^+;\mathcal K)$, and 
semigroup evolution in Hardy space of the upper half-plane $\mathcal H^2(\mathbb C^+;\mathcal K)$ or the isomorphic space 
$\mathcal H^2_+(\mathbb R;\mathcal K)$ of boundary value functions on $\mathbb R$ of functions in $\mathcal H^2(\mathbb C^+;\mathcal K)$. First we 
shall need a few more definitions. 
\par Let $\{u(t)\}_{t\in\mathbb R}$ be the family of unitary multiplicative operators 
$u(t): L^2(\mathbb R;\mathcal K)\mapsto L^2(\mathbb R;\mathcal K)$ defined by
\begin{equation*}
 [u(t)f](\sigma)=e^{-i\sigma t}f(\sigma),\qquad f\in L^2(\mathbb R;\mathcal K),\ \ \sigma\in\mathbb R,
\end{equation*}
then $\{u(t)\}_{t\in\mathbb R}$ is a continuous one parameter unitary evolution group on $L^2(\mathbb R;\mathcal K)$ and we 
have $u_+(t)=P_{\mathbb R^+}u(t)P_{\mathbb R^+}=u(t)P_{\mathbb R^+}$. Recall that we have denoted by $P_+$ the orthogonal projection of 
$L^2(\mathbb R;\mathcal K)$ on $\mathcal H^2_+(\mathbb R;\mathcal K)$. A \emph{T\"oplitz operator} ${}^{20,21,22}$ with symbol 
$u(t)$ is an operator $T_u(t):\mathcal H^2_+(\mathbb R;\mathcal K)\mapsto\mathcal H^2_+(\mathbb R;\mathcal K)$ defined by
\begin{equation*}
 T_u(t)f=P_+u(t)f,\qquad f\in\mathcal H^2_+(\mathbb R;\mathcal K)\,.
\end{equation*}
The set $\{T_u(t)\}_{t\in\mathbb R^+}$ forms a strongly continuous, contractive, one parameter semigroup on $\mathcal H^2_+(\mathbb R;\mathcal K)$ 
satisfying
\begin{equation}
\label{T_semigroup_eqn}
 \Vert T_u(t_2)f\Vert \leq \Vert T_u(t_1)f\Vert,\qquad t_2\geq t_1\geq 0,\ \ f\in\mathcal H^2_+(\mathbb R;\mathcal K)\,,
\end{equation}
and 
\begin{equation}
\label{T_to_zero_eqn}
 s-\lim_{t\to\infty} T_u(t)=0\,.
\end{equation}
Below we shall make frequent use of quasi-affine mappings. The definition of this class of maps is as follows:
\begin{definition}[quasi-affine map]
A quasi-affine map from a Hilbert space $\mathcal H_1$ into a Hilbert space $\mathcal H_0$ is a linear, injective, continuous 
mapping of $\mathcal H_1$ into a dense linear manifold in $\mathcal H_0$. If $A\in\mathscr B(\mathcal H_1)$ and 
$B\in\mathscr B(\mathcal  H_0)$ then $A$ is a quasi-affine transform of $B$ if there is a quasi-affine map $\theta:\mathcal 
H_1\mapsto \mathcal H_0$ such that $\theta A=B\theta$. \hfill$\square$
\end{definition}
Concerning quasi-affine maps we have the following two important facts (see, for example Ref. 23):
\begin{description}
\item[I)] If $\theta:\mathcal H_1\mapsto\mathcal H_0$ is a quasi-affine mapping then $\theta^*:\mathcal H_0\mapsto\mathcal H_1$
is also quasi-affine, that is, $\theta^*$ is one to one, continuous and its range is dense in $\mathcal H_1$.
\item[II)] If $\theta_1:\mathcal H_0\mapsto\mathcal H_1$ is quasi-affine and $\theta_2:\mathcal H_1\mapsto\mathcal H_2$ is
quasi-affine then $\theta_2\theta_1:\mathcal H_0\mapsto\mathcal H_2$ is quasi-affine.
\end{description}
We can now turn to the proof of Theorem 1:
\bigskip
\par{\bf Proof of Theorem 1:}
\smallskip
\par The proof of the Lyapunov propery of $M_F$ follows from an adaptation to the vector valued case of a theorem first proved in Ref. 25 and
subsequently used in the study of resonances in Refs. 24, 25, 26. We have the following proposition:
\begin{proposition}
\label{the_central_prop}
\par{\ }
\begin{description}
\item{$\alpha$)} Let $P_+\vert_{L^2(\mathbb R^+;\mathcal K)}: L^2(\mathbb R^+;\mathcal K)\mapsto\mathcal H^2_+(\mathbb R;\mathcal K)$
be the restriction of the orthogonal projection $P_+$ to the subspace $L^2(\mathbb R^+;\mathcal K)\subset L^2(\mathbb R;\mathcal K)$. Then
$P_+\vert_{L^2(\mathbb R^+;\mathcal K)}$ is a contractive quasi-affine mapping of $L^2(\mathbb R^+;\mathcal K)$ into 
$\mathcal H^2_+(\mathbb R;\mathcal K)$.
\item{$\beta$)} For $t\geq 0$ the evolution $u_+(t)$ is a quasi-affine transform of the T\"oplitz operator $T_u(t)$. For every $t\geq 0$
and $g\in L^2(\mathbb R^+;\mathcal K)$ we have
\begin{equation}
\label{basic_intertwining_rel_eqn_a}
 P_+u_+(t)g=T_u(t)P_+g,\qquad t\geq 0,\ \ g\in L^2(\mathbb R^+;\mathcal K)\,.
\end{equation}
\end{description}
\hfill$\square$
\end{proposition}
\bigskip
\par{\bf proof:}
\smallskip
\par The validity of ($\alpha$) is a result of an extension to the vector valued case of a theorem proved by Van Winter${}^{27}$ for scalar valued
functions. Taking into account the definition above of quasi-affine maps, the extension of the Van Winter theorem to vector valued functions can be
stated in a simple way:
\begin{theorem}[Van Winter theorem for vector valued functions]
\label{extended_van_winter_thm}
\par\hfill
\par Let 
$P_{\mathbb R^+}\vert_{\mathcal H^2_+(\mathbb R;\mathcal K)}:\mathcal H^2_+(\mathbb R;\mathcal K)\mapsto L^2(\mathbb R^+;\mathcal K)$ be the 
restriction of the orthogonal projection $P_{\mathbb R^+}$ to the subspace $\mathcal H^2_+(\mathbb R;\mathcal K)\subset L^2(\mathbb R^+;\mathcal K)$. 
Then $P_{\mathbb R^+}\vert_{\mathcal H^2_+(\mathbb R;\mathcal K)}$ is a contractive quasi-affine mapping of $\mathcal H^2_+(\mathbb R;\mathcal K)$ into
$L^2(\mathbb R^+;\mathcal K)$.\hfill$\square$
\end{theorem}
\par{\bf Remark:} Note that, beyond the statement made here, the original Van Winter theorem includes also explicit construction of the inverse of the 
quasi-affine map $P_{\mathbb R^+}\vert_{\mathcal H^2_+(\mathbb R)}$ (appropriate for the case of scalar valued functions) in terms of Mellin transform. 
We shall not need such explicit constructions in our discussion below.
\bigskip
\par We include here a simple proof of Theorem \ref{extended_van_winter_thm}:
\bigskip
\par{\bf Proof of Theorem \ref{extended_van_winter_thm}:}
\smallskip
\par We need to prove the following properties of the mapping 
$P_{\mathbb R^+}\vert_{\mathcal H^2_+(\mathbb R;\mathcal K)}:\mathcal H^2_+(\mathbb R;\mathcal K)\mapsto L^2(\mathbb R^+;\mathcal K)$: 
\begin{description}
\item{a)} $P_{\mathbb R^+}\vert_{\mathcal H^2_+(\mathbb R;\mathcal K)}$ is contractive,
\item{b)} $P_{\mathbb R^+}\vert_{\mathcal H^2_+(\mathbb R;\mathcal K)}$ is one to one,
\item{c)} $P_{\mathbb R^+}\vert_{\mathcal H^2_+(\mathbb R;\mathcal K)}\mathcal H^2_+(\mathbb R;\mathcal K)
 =P_{\mathbb R^+}\mathcal H^2_+(\mathbb R;\mathcal K)$ is dense in $L^2(\mathbb R^+;\mathcal K)$.
\end{description}
\par That (a) holds is an immediate consequence of the fact that $P_{\mathbb R^+}$ is a projection. The validity of (b) follows from the fact
that if $P_{\mathbb R^+}\vert_{\mathcal H^2_+(\mathbb R;\mathcal K)}$ is not one to one then there exists a function 
$f\in\mathcal H^2_+(\mathbb R;\mathcal K)$ such that $f\not=0$ and $P_{\mathbb R^+}f=0$ so that $f$ vanishes a.e. on $\mathbb R^+$. However, a Hardy 
space function $f\in\mathcal H^2_+(\mathbb R;\mathcal K)$ cannot be identically zero on any non-zero measure set in $\mathbb R$. Therefore   
$P_{\mathbb R^+}\vert_{\mathcal H^2_+(\mathbb R;\mathcal K)}$ must be one to one.
\par To prove (c) assume that $P_{\mathbb R^+}\vert_{\mathcal H^2_+(\mathbb R;\mathcal K)}\mathcal H^2_+(\mathbb R;\mathcal K)
 =P_{\mathbb R^+}\mathcal H^2_+(\mathbb R;\mathcal K)$ is not dense in $L^2(\mathbb R^+;\mathcal K)$. Then there exists a function, say 
$g\in L^2(\mathbb R^+;\mathcal K)$, such that $(g,P_{\mathbb R^+}f)_{L^2(\mathbb R^+;\mathcal K)}=0$ for all $f\in\mathcal H^2_+(\mathbb R;\mathcal K)$.
Decompose $g$ into a sum $g=g^+ +g^-$ with $g^+\in\mathcal H^2_+(\mathbb R;\mathcal K)$ and $g^-\in\mathcal H^2_-(\mathbb R;\mathcal K)$. We have
\begin{multline*}
 0=(g,P_{\mathbb R^+}\vert_{\mathcal H^2_+(\mathbb R;\mathcal K)}f)_{L^2(\mathbb R^+;\mathcal K)}
 =(g,P_{\mathbb R^+}f)_{L^2(\mathbb R^+;\mathcal K)}=(g,f)_{L^2(\mathbb R;\mathcal K)}=\\
 =(g^+ +g^-,f)_{L^2(\mathbb R;\mathcal K)}
 =(g^+,f)_{L^2(\mathbb R;\mathcal K)}=(g^+,f)_{\mathcal H^2_+(\mathbb R;\mathcal K)}\,. 
\end{multline*}
Since $f$ is arbitrary we must have $g^+=0$ and $g=g^-$. However $g$ is supported in $\mathbb R^+$ and is identically zero on $\mathbb R^-$ and $g^-$, being a 
Hardy space function, cannot identically vanish on $\mathbb R^-$. We get a contradiction.\hfill$\blacksquare$
\bigskip
\par By property (I) above of quasi-affine maps the adjoint of 
$P_{\mathbb R^+}\vert_{\mathcal H^2_+(\mathbb R;\mathcal K)}$ is also quasi-affine. For any $g\in L^2(\mathbb R^+;\mathcal K)$ and 
$f\in\mathcal H^2_+(\mathbb R;\mathcal K)$ we have,
\begin{multline}
\label{quasi_affine_proj_adjoint_eqn}
 (g,P_{\mathbb R^+}\vert_{\mathcal H^2_+(\mathbb R;\mathcal K)}f)_{L^2(\mathbb R^+;\mathcal K)}
 =(g,P_{\mathbb R^+}f)_{L^2(\mathbb R^+;\mathcal K)}=(g,P_{\mathbb R^+}f)_{L^2(\mathbb R;\mathcal K)}
 =(g,f)_{L^2(\mathbb R;\mathcal K)}=\\
 =(P_+g,f)_{L^2(\mathbb R;\mathcal K)}=(P_+\vert_{L^2(\mathbb R^+;\mathcal K)}g,f)_{\mathcal H^2_+(\mathbb R;\mathcal K)}\,.
\end{multline}
Thus, we find that $P_+\vert_{L^2(\mathbb R^+;\mathcal K)}$ is the adjoint of $P_{\mathbb R^+}\vert_{\mathcal H^2_+(\mathbb R;\mathcal K)}$ and, hence,
it is quasi-affine and contractive.
\par To prove ($\beta$) we first note that for $t\geq 0$ the Hardy space $\mathcal H^2_-(\mathbb R;\mathcal K)$ is stable under the action of 
the evolution $u(t)$, i.e., we have $u(t)\mathcal H^2_-(\mathbb R;\mathcal K)\subset\mathcal H^2_-(\mathbb R;\mathcal K)$ for all $t\geq 0$. We then 
have
\begin{multline*}
 P_+u_+(t)g=P_+u(t)P_{\mathbb R^+}g=P_+u(t)g=P_+u(t)(P_+ +P_-)g=\\
 =P_+u(t)P_+g+P_+u(t)P_-g=P_+u(t)P_+g=T_u(t)P_+g,\quad t\geq 0,\ \ g\in L^2(\mathbb R^+;\mathcal K)\,.
\end{multline*}
\par\hfill$\blacksquare$
\bigskip
\par According to ($\alpha$) in proposition \ref{the_central_prop} the restriction of the mapping $P_+$ to $L^2(\mathbb R^+;\mathcal K)$
is a contractive and quasi-affine mapping of $L^2(\mathbb R^+;\mathcal K)$ into $\mathcal H^2_+(\mathbb R^+;\mathcal K)$. Let us denote this 
restriction by $\Omega_f$. It will be convenient to first consider the mapping $P_+P_{\mathbb R^+}:L^2(\mathbb R;\mathcal K)\mapsto L^2(\mathbb R;\mathcal K)$
and then define $\Omega_f: L^2(\mathbb R^+;\mathcal K)\mapsto\mathcal H^2_+(\mathbb R;\mathcal K)$ using this operator. Thus we set 
\begin{equation*}  
 \Omega_f:=P_+P_{\mathbb R^+}\vert_{L^2(\mathbb R^+;\mathcal K)}=P_+\vert_{L^2(\mathbb R^+;\mathcal K)}
\end{equation*}
(here the subscript $f$ in $\Omega_f$ designates forward time evolution; see Eq. (\ref{basic_intertwining_rel_eqn}) below). By (I) above if $\Omega_f$ 
is contractive and quasi-affine then its adjoint 
$\Omega_f^*: \mathcal H^2_+(\mathbb R;\mathcal K)\mapsto L^2(\mathbb R^+;\mathcal K)$ is also contractive and quasi-affine. By 
Eq. (\ref{quasi_affine_proj_adjoint_eqn}) 
\begin{equation*}
 \Omega_f^*=P_{\mathbb R^+}P_+\vert_{\mathcal H^2_+(\mathbb R;\mathcal K)}=P_{\mathbb R^+}\vert_{\mathcal H^2_+(\mathbb R;\mathcal K)}\,.
\end{equation*}
Now define the Lyapunov operator $M_F: L^2(\mathbb R^+;\mathcal K)\mapsto L^2(\mathbb R^+;\mathcal K)$ by
\begin{equation*}
 M_F:=\Omega_f^*\Omega_f=P_{\mathbb R^+}P_+P_{\mathbb R^+}\vert_{L^2(\mathbb R^+;\mathcal K)}\,.
\end{equation*}
By (II) above and the fact that $\Omega_f$, $\Omega_f^*$ are quasi-affine we get that $M_F$ is a quasi-affine mapping of
$L^2(\mathbb R^+;\mathcal K)$ into $L^2(\mathbb R^+;\mathcal K)$. Therefore $M_F$ is continuous and injective and $Ran\,M_F$ is dense in 
$L^2(\mathbb R^+;\mathcal K)$. Obviously $M_F$ is symmetric and, since $\Omega_f$ and $\Omega_f^*$ are bounded, then
$Dom\, M_F=L^2(\mathbb R^+;\mathcal K)$ and we conclude that $M_F$ is self-adjoint. Since $\Omega_f$ and $\Omega_f^*$ are both contractive
then $M_F$ is contractive. 
\par Following the definition of the mappings $\Omega_f$ and $\Omega_f^*$ it is convenient to write Eq. (\ref{basic_intertwining_rel_eqn_a}) in the 
form
\begin{equation*}
 P_+P_{\mathbb R^+}u_+(t)g=T_u(t)P_+P_{\mathbb R^+}g,\qquad t\geq 0,\ \ g\in L^2(\mathbb R^+;\mathcal K)\,,
\end{equation*}
and hence
\begin{equation}
\label{basic_intertwining_rel_eqn}
 \Omega_f u_+(t)g=T_u(t)\Omega_f g,\qquad t\geq 0,\ \ g\in L^2(\mathbb R^+;\mathcal K)\,.
\end{equation}
Taking the adjoint of Eq. (\ref{basic_intertwining_rel_eqn}) we get
\begin{equation}
\label{adj_basic_intertwining_rel_eqn}
 u_+(-t)\Omega^*_f g=\Omega^*_f(T_u(t))^*g,\qquad t\geq 0,\ \ g\in\mathcal H^2_+(\mathbb R;\mathcal K)\,,
\end{equation}
and from Eqns. (\ref{basic_intertwining_rel_eqn}) and (\ref{adj_basic_intertwining_rel_eqn}) we obtain an expression for the
Heisenberg evolution of $M_F$
\begin{equation*}
 u_+(-t)M_Fu_+(t)=u_+(-t)\Omega_f^*\Omega_f u_+(t)=\Omega_f^* (T_u(t))^*T_u(t)\Omega_f\,.
\end{equation*}
For any $\psi\in L^2(\mathbb R^+;\mathcal K)$ we then get
\begin{equation*}
 (\psi,U(-t)M_FU(t)\psi)=(\psi,\Omega_f^*(T_u(t))^*T_u(t)\Omega_f\psi)
 =\Vert T_u(t)\Omega_f\psi\Vert^2,\qquad t\geq 0,\ \ \psi\in L^2(\mathbb R^+;\mathcal K)\,.
\end{equation*}
The fact that $M_F$ is a Lyapunov variable, i.e., the validity of Eqns. (\ref{M_F_lyapounov_var_eqn}) and 
(\ref{M_F_to_zero_eqn}) then follows immediately from Eqns. (\ref{T_semigroup_eqn}) and (\ref{T_to_zero_eqn}).\hfill$\blacksquare$
\bigskip
\par We proceed now to the proof of the second main result of this paper:
\bigskip
\par{\bf Proof of Theorem \ref{lambda_transform_thm}}
\smallskip
\par Since $M_F$ is a bounded positive operator its positive square root $\Lambda_F$ is well defined and unique
and we set $\Lambda_F:=M_F^{1/2}=(\Omega_f^*\Omega_f)^{1/2}$. Moreover, since 
\begin{equation*}
 M_F L^2(\mathbb R^+;\mathcal K)=\Lambda_F\Lambda_F L^2(\mathbb R^+;\mathcal K)\subseteq\Lambda_F L^2(\mathbb R^+;\mathcal K)\,,
\end{equation*}
and since $Ran\,M_F$ is dense in $L^2(\mathbb R^+;\mathcal K)$ we conclude that $Ran\,\Lambda_F$ is dense in $L^2(\mathbb R^+;\mathcal K)$. Furthermore,
since $M_F=\Lambda_F^2$ is one to one then $\Lambda_F$ must also be one to one. We can summarize the findings above by 
stating that the fact that $M_F$ is positive, one to one and quasi-affine implies the same properties for $\Lambda_F$.
Since $M_F$ is contractive and since for every $\psi\in L^2(\mathbb R^+;\mathcal K)$ we have $(\psi,M_F\,\psi)=\Vert \Lambda_F\psi\Vert^2$ we 
conclude, by using the Schwartz inequality, that $\Lambda_F$ is also contractive. 
\par Define a mapping $\tilde R: \Lambda_F L^2(\mathbb R^+;\mathcal K)\mapsto\mathcal H^2_+(\mathbb R;\mathcal K)$ by 
\begin{equation}
\label{R_tilde_def_eqn}
 \tilde R:=\Omega_f (\Omega_f^*\Omega_f)^{-1/2}=\Omega_f \Lambda_F^{-1}\,.
\end{equation}
Obviously $\tilde R$ is defined on a dense set in $L^2(\mathbb R^+;\mathcal K)$. For any $g\in \Lambda_F L^2(\mathbb R^+;\mathcal K)$ we have
\begin{multline}
\label{R_tilde_isometric_eqn}
 \Vert\tilde R g\Vert^2=(\tilde R g,\tilde R g)=(\Omega_f(\Omega_f^*\Omega_f)^{-1/2}g,\Omega_f(\Omega_f^*\Omega_f)^{-1/2}g)=\\
 =((\Omega_f^*\Omega_f)^{-1/2}g,\Omega_f^*\Omega_f(\Omega_f^*\Omega_f)^{-1/2}g)
  =((\Omega_f^*\Omega_f)^{-1/2}g,(\Omega_f^*\Omega_f)^{1/2}g)=\Vert g\Vert^2\,,
\end{multline}
hence $\tilde R$ is isometric on a dense set in $L^2(\mathbb R^+;\mathcal K)$ and can be extended to an isometric map
$R:L^2(\mathbb R^+;\mathcal K)\mapsto\mathcal H^2_+(\mathbb R;\mathcal K)$ such that 
\begin{equation*}
 R^*R=I_{L^2(\mathbb R^+;\mathcal K)}\,.
\end{equation*}
Since $R$ is an isometry $Ran\,R$ is a closed subspace of $\mathcal H^2_+(\mathbb R;\mathcal K)$. Moreover, 
$Ran R\supset\linebreak R\Lambda_F L^2(\mathbb R^+;\mathcal K)=\tilde R\Lambda_F L^2(\mathbb R^+;\mathcal K)=\Omega_f L^2(\mathbb R^+;\mathcal K)$, so 
that $Ran\,R$ contains a dense set in $\mathcal H^2_+(\mathbb R;\mathcal K)$. Therefore
$RL^2(\mathbb R^+;\mathcal K)=\mathcal H^2_+(\mathbb R;\mathcal K)$ and we conclude that 
$R:L^2(\mathbb R^+;\mathcal K)\mapsto\mathcal H^2_+(\mathbb R;\mathcal K)$ is, in fact, an isometric isomorphism.

\par From the definition of $\tilde R$ in Eq. (\ref{R_tilde_def_eqn}), or from Eq. (\ref{R_tilde_isometric_eqn}), we see that on the dense set
$\Omega_fL^2(\mathbb R^+;\mathcal K)\subset\mathcal H^2_+(\mathbb R;\mathcal K)$ we have 
\begin{equation*}
 \tilde R^* f=(\Omega_f^*\Omega_f)^{-1/2}\Omega_f^* f=\Lambda_F^{-1} \Omega_f^* f,\qquad f\in Ran\,\tilde R
\end{equation*}
and the adjoint $R^*$ of $R$ is an extension of $\tilde R^*$ to $\mathcal H^2_+(\mathbb R;\mathcal K)$. For any $g\in Ran\,\Omega_f$ we have
\begin{equation}
\label{R_tilde_star_eqn}
 \tilde R^* g=(\Omega_f^*\Omega_f)^{-1/2}\Omega_f^*g
 =(\Omega_f^*\Omega_f)^{-1/2}\Omega_f^*\Omega_f\Omega_f^{-1}g
 =(\Omega_f^*\Omega_f)^{1/2}\Omega_f^{-1}g=\Lambda_F\Omega_f^{-1}g\,.
\end{equation}
Now define
\begin{equation*}
 Z(t):=\Lambda_F U(t)\Lambda_F^{-1},\qquad t\geq 0\,.
\end{equation*}
Obviously, $Z(t)$ is well defined on $Ran\,\Lambda_F$ for any $t\geq 0$. Moreover, using the definition of $\tilde R$ from 
Eq. (\ref{R_tilde_def_eqn}) and Eqns. (\ref{R_tilde_star_eqn}), (\ref{basic_intertwining_rel_eqn}) we get
\begin{multline*}
 RZ(t)R^* g=\Omega_f\Lambda_F^{-1} Z(t)\Lambda_F\Omega_f^{-1} g
 =[\Omega_f\Lambda_F^{-1}]\,[\Lambda_F U(t)\Lambda_F^{-1}]\,[\Lambda_F\Omega_f^{-1}]g=\\
 =\Omega_f U(t)\Omega_f^{-1}g=T_u(t)g,\quad t\geq 0,\ \ g\in Ran\,\Omega_f\subset\mathcal H^2_+(\mathbb R;\mathcal K)\,.
\end{multline*}
Then on the dense subset $R^*\Omega_F L^2(\mathbb R^+;\mathcal K)\subset L^2(\mathbb R^+;\mathcal K)$ we have $Z(t)=R^*T_u(t)R$ and since $R$ and
$T_u(t)$ are bounded we are able by continuity to extend the domain of definition of $Z(t)$ to all of $L^2(\mathbb R^+;\mathcal K)$ and obtain
\begin{equation}
\label{Z_unitary_transform_eqn}
 RZ(t)R^*=T_u(t),\quad Z(t)=R^*T_u(t)R,\ \ t\geq 0\,.
\end{equation}
From the fact that $R$ is an isometric isomorphism, the fact that $\{T_u(t)\}_{t\in\mathbb R^+}$ is a continuous, strongly contractive, one parameter
semigroup and Eqns. (\ref{T_semigroup_eqn}), (\ref{T_to_zero_eqn}) we conclude that $\{Z(t)\}_{t\in\mathbb R^+}$ is a
continuous, strongly contractive, one parameter semigroup and Eqns. (\ref{Z_semigroup_prop_eqn}) and 
(\ref{Z_intertwining_eqn}) hold. \hfill$\blacksquare$
\par\bigskip\ 
\par{\bf Proof of Theorem \ref{T_observable_thm}:}
\smallskip
\par The results stated in Theorem \ref{T_observable_thm} are consequences of the following lemma:
\begin{lemma}
\label{hardy_semigroup_projections_lemma}
For every $t\leq 0$ the operator $T_u(t):\mathcal H^2_+(\mathbb R;\mathcal K)\mapsto\mathcal H^2_+(\mathbb R;\mathcal K)$ is isometric and we have 
\begin{equation}
\label{T_star_range_eqn}
 Ran\,(T_u(t))^*=(Ker\,T_u(t))^\perp
\end{equation}
and
\begin{equation*}
 T_u(t)(T_u(t))^*=I_{\mathcal H^2_+(\mathbb R;\mathcal K)},\quad t\geq 0\,.
\end{equation*}
Furthermore, if $\hat P_{t]}:\mathcal H^2_+(\mathbb R;\mathcal K)\mapsto\mathcal H^2_+(\mathbb R;\mathcal K)$ is the orthogonal projection on 
$Ker\,T_u(t)$ and $\hat P_{[t}$ is the orthogonal projection on $(Ker\,T_u(t) )^\perp$ then 
\begin{equation*}
 \hat P_{t]}=[T_u(t),(T_u(t))^*],\quad t\geq 0
\end{equation*}
and
\begin{equation*}
 \hat P_{[t}=(T_u(t))^*T_u(t),\quad t\geq 0\,.
\end{equation*} 
Moreover, we have
\begin{equation}
\label{P_t_right_product_eqn}
 \hat P_{t_1]}\hat P_{t_2]}=\hat P_{t_1]},\ \ t_2\geq t_1\geq 0,\qquad Ran\,\hat P_{t_1]}\subset Ran\,\hat P_{t_2]},\quad t_2>t_1
\end{equation}
and 
\begin{equation*}
 \hat P_{0]}=0,\qquad \lim_{t\to\infty} \hat P_{t]}=I_{\mathcal H^2_+(\mathbb R;\mathcal K)}\,.
\end{equation*}
\hfill$\square$
\end{lemma}
\par{\bf Proof of Lemma \ref{hardy_semigroup_projections_lemma}:}
\smallskip
\par Recall that $T_u(t)f=P_+u(t)f$ for $t\geq 0$ and $f\in\mathcal H^2_+(\mathbb R;\mathcal K)$. Since $\mathcal H^2_+(\mathbb R;\mathcal K)$ is 
stable under $u^*(t)=u(-t)$ for $t\geq 0$, i.e., $u(-t)\mathcal H^2_+(\mathbb R;\mathcal K)\subset\mathcal H^2_+(\mathbb R;\mathcal K)$ (as one can 
see, for example, by using the Paley-Wiener theorem${}^{28}$), we find that for any $f,g\in\mathcal H^2_+(\mathbb R;\mathcal K)$ we have
\begin{equation*}
 (g,T_u(t)f)=(g,P_+u(t)f)=(u(-t)g,f)=(P_+u(-t)g,f)=(T_{u^*}(t)g,f)=((T_u(t))^*g,f)\,.
\end{equation*}
Therefore
\begin{equation}
\label{T_star_expression_eqn}
 (T_u(t))^*g=u(-t)g,\quad t\geq 0,\ \ g\in\mathcal H^2_+(\mathbb R;\mathcal K)\,.
\end{equation}
Since $u(-t)$ is unitary on $L^2(\mathbb R;\mathcal K)$ Eq. (\ref{T_star_expression_eqn}) implies that $(T_u(t))^*$ is isometric
on $\mathcal H^2_+(\mathbb R;\mathcal K)$. The same equation implies also that
\begin{equation}
\label{T_T_star_unit_eqn}
 (T_u(t)(T_u(t))^*f=P_+u(t)u(-t)f=P_+f=f,\quad t\geq 0,\ \ f\in\mathcal H^2_+(\mathbb R;\mathcal K)\,.
\end{equation}
Consider now the operator $A(t):=T_u(t))^*T_u(t)$ for $t\geq 0$. Since $Dom\,T_u(t)=\mathcal H^2_+(\mathbb R;\mathcal K)$ we have
that $A(t)$ is self-adjoint. In addition Eq. (\ref{T_T_star_unit_eqn}) implies that 
\begin{equation*}
 (A(t))^2=[(T_u(t))^*T_u(t)][(T_u(t))^*T_u(t)]=(T_u(t))^*T_u(t)=A(t),\quad t\geq 0\,,
\end{equation*} 
so that $A(t)$ is an orthogonal projection in $\mathcal H^2_+(\mathbb R;\mathcal K)$. Of course, for any $u\in Ker\,T_u(t)$ we have
$A(t)u=0$, hence $Ran\,A(t)\subseteq (Ker\,T_u(t))^\perp$. Assume that there is some 
$v\in(Ran\,A(t))^\perp\cap(Ker\,T_u(t))^\perp$ with $v\not= 0$. Then we must have $A(t)v=0$, but since $T_u(t)v\not=0$
and since $(T_u(t))^*$ is an isometry we obtain a contradiction. Therefore $Ran\,A(t)=(Ker\,T_u(t))^\perp$ and
$\hat P_{[t}=A(t)=(T_u(t))^*T_u(t)$. Taking into account Eq. (\ref{T_T_star_unit_eqn}) we obtain also
$\hat P_{t]}=I-\hat P_{[t}=[T_u(t),(T_u(t))^*]$.
\par To prove Eq. (\ref{T_star_range_eqn}) we note that since $(T_u(t))^*$ is isometric its range is a close subspace of 
$\mathcal H^2_+(\mathbb R;\mathcal K)$ and, moreover, $(Ran\,(T_u(t))^*)^\perp\supseteq Ker\,T_u(t)$. This is a result of the fact that if
$u\in Ker\,T_u(t)$ then $(u,(T_u(t))^*v)=(T_u(t)u,v)=0$, $\forall v\in\mathcal H^2_+(\mathbb R;\mathcal K)$. On the other hand, if $u$ is orthogonal to 
$Ran\,(T_u(t))^*$ i.e., $u$ is such that $(u,(T_u(t))^*v)=0$, $\forall v\in\mathcal H^2_+(\mathbb R;\mathcal K)$ then 
$(T_u(t)u,v)=0$, $\forall v\in\mathcal H^2_+(\mathbb R;\mathcal K)$ so that $u\in Ker\,T_u(t)$ and we get that 
$(Ran\,(T_u(t))^*)^\perp\subseteq Ker\,T_u(t)$.
\par In order to verify the validity of the first equality in Eq. (\ref{P_t_right_product_eqn}) we use the semigroup property of 
$\{T_u(t)\}_{t\in\mathbb R^+}$ and Eq. (\ref{T_T_star_unit_eqn}). For $t_1\leq t_2$ we get
\begin{multline*}
 \hat P_{t_1]}\hat P_{t_2]}=(I-\hat P_{[t_1})(I-\hat P_{[t_2})
 = I-\hat P_{[t_1}-\hat P_{[t_2}+(T_u(t_1))^*T_u(t_1)(T_u(t_2))^*T_u(t_2)=\\
 =I-\hat P_{[t_1}-\hat P_{[t_2}+(T_u(t_1))^*(T_u(t_2-t_1))^*T_u(t_2)
 =I-\hat P_{[t_1}-\hat P_{[t_2}+(T_u(t_2))^*T_u(t_2)=\\
  =I-\hat P_{[t_1}-\hat P_{[t_2}+\hat P_{[t_2}=\hat P_{t_1]}\,.
\end{multline*}
We need to show also that $Ker\,T_u(t_1)\subset Ker\,T_u(t_2)$ for $t_2>t_1$. Note that since for $t_2>t_1$ we have
$T_u(t_2)=T_u(t_2-t_1)T_u(t_1)$ and since $(T_u(t))^*$ is isometric on $\mathcal H^2_+(\mathbb R;\mathcal K)$ then it is enough
to show that $Ker\,T_u(t)\not=\{0\}$ for every $t>0$. If this condition is true and if $f\in Ker\,T_u(t_2-t_1)$
we just set $g=(T_u(t_1))^*f$ and we get that 
\begin{equation*}
 T_u(t_1)g=T_u(t_1)(T_u(t_1))^*f=f
\end{equation*}
and
\begin{equation*}
 T_u(t_2)g=T_u(t_2)(T_u(t_1))^*f=T_u(t_2-t_1)f=0\,.
\end{equation*}
In order to show that $Ker\,T_u(t)\not=\{0\}$ for every $t>0$ we exhibit a state belonging to this kernel. Indeed one may easily
check that for a complex constant $\mu$ such that $Im\,\mu<0$, any vector $v\in\mathcal K$ and for $t_0>0$ the function
\begin{equation*}
 f(\sigma)=\frac{1}{\sigma-\mu}\left[1-e^{i\sigma t_0}e^{-i\mu t_0}\right]v,\quad \sigma\in\mathbb R
\end{equation*}
is such that $f\in Ker\,T_u(t)\subset\mathcal H^2_+(\mathbb R;\mathcal K)$ for every $t\geq t_0>0$.
\par Finally, it is immediate that $\hat P_{0]}=0$ and, moreover, since for every $f\in\mathcal H^2_+(\mathbb R;\mathcal K)$ we have
$\Vert \hat P_{[t} f\Vert^2=(f,\hat P_{[t} f)=(f,(T_u(t))^*T_u(t)f)=\Vert T_u(t) f\Vert^2$ then $s-\lim_{t\to\infty}\hat P_{[t}=0$ and
hence $s-\lim_ {t\to\infty}\hat P_{t]}=I_{\mathcal H^2_+(\mathbb R;\mathcal K)}$.\hfill$\blacksquare$
\par\bigskip\
\par For $t\geq 0$ define $P_{t]}:=R^*\hat P_{t]}R$ and $P_{[t}:=R^*\hat P_{[t}R=I_{L^2(\mathbb R^+;\mathcal K)}-P_{t]}$. Combining Lemma 
\ref{hardy_semigroup_projections_lemma} and Eq. (\ref{Z_unitary_transform_eqn}), and taking into account the fact that the mapping $R$ is an isometric 
isomorphism, we conclude that there exists families $\{P_{t]}\}_{t\in\mathbb R^+}$, $\{P_{[t}\}_{t\in\mathbb R^+}$,  of orthogonal projections in 
$L^2(\mathbb R^+;\mathcal K)$ such that $P_{t]}+P_{[t}=I_{L^2(\mathbb R^+;\mathcal K)}$ and 
\begin{equation*}
 Ran\,P_{t]}=Ker\,Z(t),\quad Ran\,P_{[t}=(Ker\,Z(t))^\perp,\quad t\geq 0\,,
\end{equation*}
\begin{equation*}
 P_{t]}=[Z(t),Z^*(t)],\quad t\geq 0\,,
\end{equation*}
\begin{equation*}
 P_{[t}=Z^*(t)Z(t),\quad t\geq 0\,,
\end{equation*} 
\begin{equation}
\label{P_Z_right_product_eqn}
 P_{t_1]} P_{t_2]}=P_{t_1]},\ \ t_2\geq t_1\geq 0,\qquad Ran\, P_{t_1]}\subset Ran\, P_{t_2]},\quad t_2>t_1
\end{equation}
and 
\begin{equation}
\label{P_Z_limits_eqn}
 P_{0]}=0,\qquad \lim_{t\to\infty} P_{t]}=I_{L^2(\mathbb R^+;\mathcal K)}\,.
\end{equation}
In addition we have
\begin{equation*}
 Ran\,(Z^*(t))=(Ker\,Z(t))^\perp
\end{equation*}
and
\begin{equation*}
 Z(t)Z^*(t)=I_{L^2(\mathbb R^+;\mathcal K)},\quad t\geq 0\,.
\end{equation*}
Eqns. (\ref{P_Z_right_product_eqn}), (\ref{P_Z_limits_eqn}) imply that it is possible to construct from the family
$\{P_{t]}\}_{t\in\mathbb R^+}$ of orthogonal projections a spectral family of a corresponding self-adjoint operator. First define for
intervals
\begin{equation*}
\mu_T(A)=\begin{cases}
            P_{b]}-P_{a]},                 & A=(a,b]\,.\\
            P_{b]}-P_{(a-0^+)]},         & A=[a,b]\,,\\
            P_{(b-0^+)]}-P_{a]},         & A=(a,b)\,,\\
            P_{(b-0^+)]}-P_{(a-0^+)]}, & A=[a,b)\,,\\
             \end{cases}
\end{equation*}
where $b>a> 0$ (and with $P_{(a-0^+)]}$ replaced by $P_{0]}$ for $a=0$), and then extend $\mu_T$ to the Borel $\sigma$-algebra 
of $\mathbb R^+$. Following the definition of the spectral measure 
$\mu_T: \mathscr B(L^2(\mathbb R^+;\mathcal K))\mapsto \mathscr P(L^2(\mathbb R^+;\mathcal K))$ we subsequently are able to define a
self-adjoint operator $T:\mathscr D(T)\mapsto L^2(\mathbb R^+;\mathcal K)$ via
\begin{equation*}
 T:=\int_0^\infty t\,d\mu_T(t)\,.
\end{equation*}
By construction it is immediate that $T$ has the properties listed in Theorem \ref{T_observable_thm}. For example, we have
\begin{equation*}
 \mu([0,t])L^2(\mathbb R^+;\mathcal K)=(P_{t]}-P_{0]})L^2(\mathbb R^+;\mathcal K)=P_{t]}\mathcal H=[Z^*(t),Z(t)]L^2(\mathbb R^+;\mathcal K)
\end{equation*}
and
\begin{multline*}
 \mu([t,\infty)L^2(\mathbb R^+;\mathcal K)=\lim_{t'\to\infty}(P_{t']}-P_{t]})L^2(\mathbb R^+;\mathcal K)
 =(I_{L^2(\mathbb R^+;\mathcal K)}-P_{t]})L^2(\mathbb R^+;\mathcal K)=\\
 =P_{[t}L^2(\mathbb R^+;\mathcal K)=Z^*(t)Z(t)L^2(\mathbb R^+;\mathcal K)\,.
\end{multline*}
\hfill$\blacksquare$
\par\smallskip\ 
\par This concludes the proofs of the three main results of this paper.
\section{Summary}
\label{summary}
\par Consider a quantum mechanical problem with corresponding Hilbert space $\mathcal H$ and Hamiltonian $H$ satisfying 
$\sigma_{ac}(H)=\mathbb R^+$ and such that the absolutely continuous subspace $\mathcal H_{ac}\subseteq\mathcal H$ is represented in the spectral 
representation for $H$ (the energy representation) in terms of a function space $L^2(\mathbb R^+;\mathcal K)$ with $\mathcal K$ a seperable Hilbert 
space. It was shown that it is then possible to construct a Lyapounov variable 
for the Schr\"odinger evolution $U(t)=\exp(-i Ht)$, $t\geq 0$ generated by $H$. The method of proof of the existence of the 
Lyapounov variable is constructive and an explicit expression for such an operator is given in the form of 
Eq. (\ref{M_F_expression_eqn}). Moreover,
it is shown that a $\Lambda$-transformation to an irreversible representation of the dynamics can be defined in this case.
Finally, it is demonstrated that the irreversible representation of the dynamics is a natural representation for the flow of time in the system in the 
sense that there exists a positive, semibounded operator $T$ on $L^2(\mathbb R^+;\mathcal K)$ such that if $\mu_T$ is the spectral projection valued 
measure of $T$ then for each $t\geq 0$ the spectral projections $P_{t]}=\mu_T([0,t))$ and 
$P_{[t}=(I_{L^2(\mathbb R^+;\mathcal K)}-P_{t]})=\mu_T([t,\infty))$ split the Hilbert space $L^2(\mathbb R^+;\mathcal K)$ into the direct sum of a 
past subspace $L^2_{t]}(\mathbb R^+;\mathcal K)$ and a future subspace $L^2_{[t}(\mathbb R^+;\mathcal K)$
\begin{equation*}
\begin{split}
 L^2(\mathbb R^+;\mathcal K)&=L^2_{t]}(\mathbb R^+;\mathcal K)\oplus L^2_{[t}(\mathbb R^+;\mathcal K),\\
 L^2_{t]}(\mathbb R^+;\mathcal K)&=P_{t]}L^2(\mathbb R^+;\mathcal K),\quad L^2_{[t}(\mathbb R^+;\mathcal K)=P_{[t}L^2(\mathbb R^+;\mathcal K)
 ,\ \ t\geq 0
 \end{split}
\end{equation*}
such that, as its name suggests, the past subspace $L^2_{t]}(\mathbb R^+;\mathcal K)$ at time $t\geq 0$ does not enter into the calculation of
matrix elements of observables for all times $t' > t\geq 0$, i.e., at time $t$ it already belongs to the past. Put 
differently, in the irreversible representation the operator $T$ provides us with a super selection rule separating past and 
future as observables of the system cannot connect the past subspace to the future subspace and matrix elements and 
expectation values for $t'>t>0$ are, in fact, calculated in the future subspace $L^2_{[t}(\mathbb R^+;\mathcal K)$.
\par\bigskip\ 
\par\centerline{\Large{\bf Acknowledgements}}
\smallskip
\par This research was supported by the Israel Science Foundation (Grant No. 1169/06 and Grant No. 1282/05). The author 
wishes to thank Prof. I.E. Antoniou for discussions motivating the present work.
\vskip 1cm
\par\noindent {\bf\Large References}
\bigskip
\begin{description}
\item{${}^1$} P. Busch, in \emph{ Time in Quantum Mechanics - Vol. 1, 2nd ed.}, edited by J. G. Muga, R. Sala Mayato and I. L. Egusquiza (Springer, 
2007).
\item{${}^2$} W. Pauli, in \emph{Handbuch der Physik, 2nd ed., Vol 24}, edited by F. Geiger and K. Scheel (Springer-Verlag, Berlin, 1933). Translated 
to english in \emph{General Principles of Quantum Mechanics} (Springer-Verlag, New-York, 1980).
\item{${}^3$} A. Galindo, Lett. Math. Phys. {\bf 8}, 495 (1984).
\item{${}^4$} J.C. Garrison and J. Wong, J. Math. Phys. {\bf 11}, 2242 (1970).
\item{${}^5$} E.A. Galapon, Proc. R. Soc. Lond. A {\bf 458}, 2671 (2002); E.A. Galapon, Proc. R. Soc. Lond. A {\bf 458}, 451
 (2002).
\item{${}^6$} A. Holevo, \emph{Probabilistic and Statistical Aspects of Quantum Theory} (North Holland, Amsterdam 1980). 
\item{${}^7$} R. Werner, J. Math. Phys. {\bf 27}, 793 (1986).
\item{${}^8$} R. Giannitrapani, Int. J. Theo. Phys. {\bf 36}, 1575 (1997).
\item{${}^9$} I.L. Egusquiza, J.G. Muga and A.D. Baute, in \emph{Time in Quantum Mechanics - Vol. 1, 2nd ed.}, edited by J. G. Muga, R. Sala Mayato and I. L. Egusquiza (Springer, 2007).
\item{${}^{10}$} P. Busch, M. Grabowski and P.J. Lahti, \emph{Operational Quantum Physics} (Springer-Verlag, Berlin 1995).
\item{${}^{11}$} P. Koosis, \emph{Introduction to $H^p$ spaces} (Cambridge: Cambridge University Press, 1980).
\item{${}^{12}$} P.L. Duren, \emph{Theory of $H^p$ spaces} (New York: Academic, 1970).
\item{${}^{13}$} K. Hoffman, \emph{Banach spaces of analytic functions} (Englewood cliffs NJ: Prentice-Hall, 1962).
\item{${}^{14}$} B. Misra, Proc. Natl. Acad. Sci. {\bf 75}, 1627 (1978).
\item{${}^{15}$} B. Misra, I. Prigogine and M. Courbage M, Proc. Natl. Acad. Sci. {\bf 76}, 4768 (1979).
\item{${}^{16}$} B. Misra, I. Prigogine and M. Courbage M, Physica A {\bf 98A}, 1 (1979).
\item{${}^{17}$} M. Courbage, Lett. Math. Phys. {\bf 4}, 425 (1980).
\item{${}^{18}$} I. Prigogine, \emph{From being to becoming} (New York: W.H. Freeman and company, 1980).
\item{${}^{19}$} I.E. Antoniou and B. Misra, J. Phys. A {\bf 24}, 2723 (1991).
%
%
\item{${}^{20}$} M. Rosenblum and J. Rovnyak \emph{Hardy classes and operator theory} (New York: Oxford University Press 1985).
\item{${}^{21}$} N.K. Nikol'ski\u{i}, \emph{Treatise on the shift operator: spectral function theory} (Springer, New York 1986).
\item{${}^{22}$} N.K. Nikol'ski\u{i}, \emph{Operators, functions and systems, an easy reading, Vol I} (American Mathematical Society, Providence 2002).
\item{${}^{23}$} B. Sz.-Nagy and C. Foias, \emph{Harmonic analysis of operators in Hilbert space} (North-Holland, Amsterdam 1970).
\item{${}^{24}$} Y. Strauss, L.P. Horwitz and A. Volovick, J. Math. Phys. {\bf 47}, 123505(1-19) (2006).
\item{${}^{25}$} Y. Strauss, J. Math. Phys. {\bf 46}, 032104(1-25) (2005).
\item{${}^{26}$} Y. Strauss, J. Math. Phys. {\bf 46}, 102109(1-12) (2005).
\item{${}^{27}$} C. Van-Winter, Trans. Am. Math. Soc. {\bf 162}, 103 (1971).
\item{${}^{28}$} R.E.A.C. Paley and N. Wiener, \emph{Fourier transforms in the complex domain} (New York: American 
Mathematical Society Colloq. Pub., Vol. 19, 1934)
\end{description}
\bigskip

\end{document}